# IRTF/SPEX Observations of the Unusual Kepler Lightcurve System KIC8462852

C.M. Lisse[1], M.L. Sitko[2], M. Marengo[3]

[1] JHU-APL, 11100 Johns Hopkins Road, Laurel, MD 20723 carey.lisse@jhuapl.edu.

[2] Department of Physics, University of Cincinnati, Cincinnati, OH 45221-0011

[3] Department of Physics and Astronomy, 12 Physics Hall, Iowa State University, Ames, IA 50010

10 Pages, 2 Figures

# Abstract

We have utilized the NASA/IRTF 3m SpeX instrument's high resolution spectral mode (Rayner *et al.* 2003) to observe and characterize the near-infrared flux emanating from the unusual Kepler lightcurve system KIC8462852. By comparing the resulting $0.8 - 4.2$ um spectrum to a mesh of model photospheric spectra, the 6 emission line analysis of the Rayner *et al.* 2009 catalogue, and the 25 system collection of debris disks we have observed to date using SpeX under the Near InfraRed Debris disk Survey (NIRDS; Lisse *et al.* 2016), we have been able to additionally characterize the system. Within the errors of our measurements, this star looks like a normal solar abundance main sequence F1V to F3V dwarf star without any obvious traces of significant circumstellar dust or gas. Using Connelley & Greene's (2014) emission measures, we also see no evidence of significant ongoing accretion onto the star nor any stellar outflow away from it. Our results are inconsistent with large amounts of static close-in obscuring material or the unusual behavior of a YSO system, but are consistent with the favored episodic models of a Gyr old stellar system favored by Boyajian *et al.* (2015). We speculate that KIC8462852, like the ~1.4 Gyr old F2V system η Corvi (Wyatt *et al.* 2005, Chen *et al.* 2006, Lisse *et al.* 2012), is undergoing a Late Heavy Bombardment, but is only in its very early stages.

# 1. Introduction

KIC 8462852, also known as TYC 3162−665−1 or 2MASS J20061546+4427248, is a V ~12, K ~10.5 mag star located in the starfield of the Kepler space telescope primary mission (Borucki *et al.* 2010). This star was identified serendipitously by the Planet Hunters project (Fischer *et al.* 2012) for its unusual light-curve, characterized by deep dimming (down to below ~ 20% of the stellar flux) lasting between 5 to 80 days, and with an irregular cadence and unusual profile (Boyajian *et al.* 2015, B15 hereafter). Since its discovery, KIC 8462852 has been the subject of intense multi-wavelength monitoring, and has spurred numerous speculations about the nature of the bodies, or structures, responsible for the dimming of its visible flux (see e.g. Wright *et al.* 2015).

KIC 8462852 was carefully characterized in B15. High resolution (R ~ 47000) spectroscopic observations obtained with the FIbre-fed Echelle Spectrograph (FIES) spectrograph at the Nordic Optical Telescope in La Palma, Spain, revealed that KIC 8462852 is a main sequence star with an effective temperature $T_{eff}$ = 6750±120 K, log g = 4.0±0.2 and solar metallicity, consistent with a F3 IV/V star. The spectral energy distribution of the source, obtained combining ground-based BV (RI)c and 2MASS (Skrutskie *et al.* 2006) $JHK_s$ photometry, with space-based NUV Galex (Morrissey *et al.* 2007) and mid-IR WISE (Wright *et al.* 2010) data, is also consistent with the spectroscopic identification of the source. Careful fitting of the star's spectral energy distribution (SED) with a stellar atmosphere model revealed that the source is located at a distance of 454 pc, with a reddening of E(B − V ) = 0.11 mag.

Interest in KIC 8462852 is related to the very unusual dips in its Kepler time series photometry. One event near JD 2455626 (2011 March 5) lasting for ~ 3 days, and a series of events starting from JD 2456343 (2013 February 19) lasting for ~ 60 days, stand out for their unusual shape, lack of periodicity and depth. A search through the Kepler database for targets with similar dips turned out empty-handed. Realistic scenarios for the phenomena observed for this star are discussed at length in B15. They are related to the episodic occultation of the star by a circumstellar dust clump, either produced in the aftermath of a catastrophic collision in the system's asteroid belt or a giant impact in the system, or associated to a population of dust-enshrouded planetisimals, or produced by the breakout of a family of comets.

The parameter space for evaluating all these scenarios is constrained by the lack of any large excess above the expected stellar photospheric emission in the mid-IR from WISE (observing before the dimming events, Boyajian *et al.* 2015) and Spitzer/IRAC (observing after the Kepler dimming events, Marengo *et al.* 2015). However, only moderate constraints have been placed to date on the possibility of very close-in, hot (> 1000 K) circumstellar dust and gas within a fraction of an AU of the primary star by the stellar model fits to broad-band infrared photometry (Boyajian et al. 2015, Marengo *et al.* 2015). To broad-band photometry, a hot dust excess would appear as an increased Rayleigh-Jeans flux from the stellar photoshere, potentially masking any excess in the mid-IR as well. Such hot dust or gas, created for example by a roasting planet like WASP-12b, or by a remnant primordial disk (e.g. Sitko *et al.* 2012), could intercept a large solid angle of starlight and create the observed light curve minima, while re-irradiating the bulk of its energy in the near-IR, invisible to the WISE and Spitzer spacecraft.

Having recently completed a 25 system survey of northern debris disks using the NASA/IRTF 3m's SPEX high resolution spectrometer at 0.8 – 5.0 um (Rayner *et al.* 2003, 2009; Vacca *et al.* 2003, 2004; Cushing *et al.* 2004), including systems showing NIR excesses at the 1-100% level above the photosphere, we were well placed to observe and characterize the system in the near-infrared. We thus requested and received Director's Discretionary Time from the IRTF to observe the system on 31 October 2015 in the same fashion as for our 25 Near InfraRed Debris disk Survey (NIRDS; Lisse *et al.* 2016) targets, and we report the results of our observations here.

## 2. Observations

We observed the KIC 8462852 system from 04:05 – 06:35 31 Oct 2015 UT (at 18:05 – 20:35 30 Oct 2015 HST time, or evening twilight) from the NASA/IRTF 3m on the summit of Mauna Kea, HI. The SpeX instrument provides R = 2000 to 2500 resolution observations from 0.8 – 5.0 um in two orders, termed SXD (for "short cross-dispersed") and LXD (for "long cross-dispersed") when configured for use using an 0.3" slit and the standard optics chain (Rayner *et al.* 2003; 2009). Our observational setup was identical to that used for our previous NIRDS debris disk studies, with two exceptions designed to maximize our chances of detecting this unknown, relatively faint source in the 2.5 total hours we had to conduct our observations (most NIRDS objects were of [K] = 2 to 8, while KIC 8462852 has a 2MASS K-magnitude of 10.54; Lisse *et al.* 2016). We utilized an 0.8" wide slit, and we doubled up on the spectral overlap between our SXD (0.8 – 2.55 um) and LXD orders by utilizing the "short LXD" 1.67 – 4.2 um mode. The result was to produce spectra with 8/3 the limiting sensitivity per on-target second of our normal survey, with spectral grasp of 0.8 – 4.2 um and R= 800 - 900. The nearby A0V star used in our SPEXTOOLS data reduction as a calibration standard (Vacca *et al.* 2004) was the [K] = 6.4 star HD 192538. We observed the standard in SXD mode with a total on-target integration time of 480 sec, and in LXD mode for a total time of 960 sec. We observed KIC8462852 in SXD mode for a total time of 2400 sec and in LXD mode for a total time of 2340 sec. The instrument behavior was nominal and the summit weather was good, with clear skies, a relative humidity of 46% and outside temperature of 4.6$^o$ C, a low CFHT Skyprobe camera attenuation value of 0, and a seeing of ~1". All objects were observed in ABBA nod mode to remove telescope and sky backgrounds. We observed KIC8462852 after meridian transit, and the position angle of the slit on the sky was continually adjusted from 160$^o$ to 134$^o$ to be perpendicular to the horizon, so the 3 magnitudes fainter M-star companion reported by Boyajian *et al.* 2015 at 2" to the East (PA 0$^o$) would not have contributed significantly to our measurements.

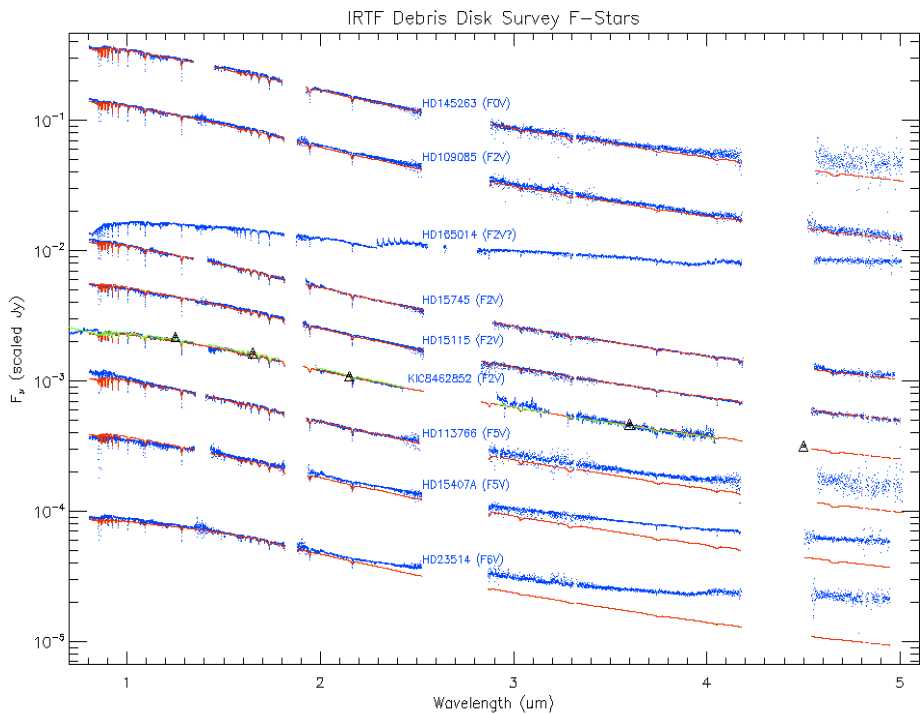

**Figure 1 – The as-observed SpeX KIC8462852 spectrum** in comparison to 8 F-stars of the NIRDS program with reported optical or IR debris disks. Blue points are the as-observed SpeX R=1000 to 2000 spectra, and red curves are solar abundance model Kurucz spectra reddened to the same amount of extinction. KIC 8462852's spectrum, 6$^{th}$ from the top, with E(B-V) = 0.11, is consistent with purely stellar photospheric emission within the uncertainties of our measurements (~ 1% at 0.8 – 2.55 um, and ~3% from 3.0 to 4.2 um). Also plotted on top of the KIC8462852 spectrum are the JHK/3.6/4.5 photometry (triangles) and the best-fit Phoenix photospheric model (green) from Marengo et al. (2015)

## 3. Results

KIC's spectrum is consistent with purely stellar photospheric emission, within the uncertainties of the measurement (Figure 1). By comparison to the spectra obtained for 8 F-stars with reported debris disks as part of the NIRDS program, KIC 8462852 clearly does not show the markedly flattened, non-photospheric hot dust continuum + ~2.0 um CO emission lines of YSO HD 165014, a system which had previously been reported in the literature as a main sequence star (e.g. Fujiwara *et al.* 2010). Nor does it show evidence for the warm dust continuum excesses rising towards long wavelengths of the 10 – 200 Myr old systems HD 145263, HD 113766, HD 15407A, or HD 23514, all with known strong 10 and 20 um IRAS and Spitzer excesses (Chen *et al.* 2006). Instead, KIC8462852's spectrum is consistent with that of the purely photospheric systems HD 15745 and HD 15115. It may have a small fractional excess hidden in the noise of our short LXD measurement, but it is clearly less than the ~5% at 3.6 um and 2.5% at 4.5 um we have found for the 1.4 Gyr old HD 109085 (η Corvi) F2V system (Lisse *et al.* 2012, Lisse, Marengo, *et al.* 2016).

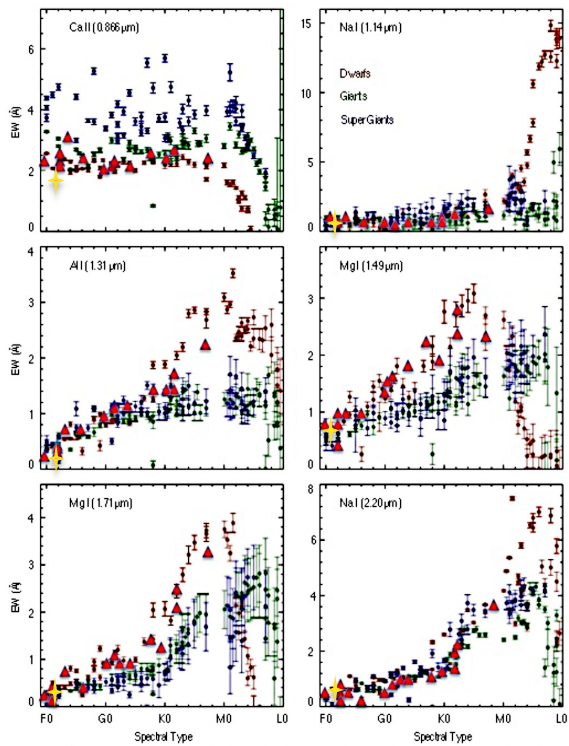

**Figure 2 – Spectral classification of KIC 8462852** (yellow star) using the SPEX 6-absorption line trends of Rayner *et al.* (2009). The Rayner *et al.* library stars, dwarfs through supergiants, are shown using the small dark blue, green, and red symbols, while our NIRDS debris disks stars are denoted by the large bright red triangles.

We have performed an independent spectral classification of KIC 8462852 using the 6-absorption line emission width (EW) trends of Rayner *et al.* (2009). These trends, based on the SPEX spectra of ~200 main sequence catalogued in the IRTF spectral library (http://irtfweb.ifa.hawaii.edu/~SPEX/IRTF_ Spectral _Library/), produced consistent determinations with respect to Kurucz model fitting of our NIRDS systems (Lisse *et al.* 2016; Figure 2). For KIC 8462852, shown as a yellow star in Fig. 2, we find EW's of 1.6 (Ca II 0.866 um), 0.3 (Na I 1.14 μm), 0.0 (Al I 1.313 um), 0.65 (MgI 1.485 um), 0.25 (MgI 1.711 um), and 0.30 (Na I 2.206 um), leading to a stellar classification of F1V to F2V. The rather low EW for CaII rules out a sub-giant classification. The spectrum of KIC 8462852 also matches well to that of a canonical main sequence F2V Kurucz spectrum with $T_{eff}$ = 6945 K, log g = +4.0, and log [Fe/H] = 0.0 and to a Phoenix next-generation photosopheric model with $T_{eff}$ = 6890 K, log g = +4.0, and log [Fe/H] = 0.0 used by Marengo et al. (2015) to analyze their 2MASS and Spitzer photometry of the system.

Following Connelly & Greene (2010, 2014), we have also examined the KIC 8462852 data for HBrγ and CO emission lines due to accretion of hot gas onto the central primary, and for HeI 1.083 μm, Fe II 1.256/1.644 μm, and H2 S(1) 1-0 2.121 μm emission lines due to strong wind outflow, and find none detectable within the noise of the measurement.

## 4. Discussion & Conclusions

As mentioned in the results section, there is a small mismatch between the stellar classification results obtained by a Rayner 6-line analysis and Kurucz model matching. The 6-line classification produces an F1V to F2V, with a very low absorption measure (reminiscent of the depression we saw for HR8789A in our NIRDS survey, a known λ Boo star, vs. our other solar abundance objects). The best solar abundance Kurucz model match puts the star somewhere between F2V and F3V due to the flattening or turnover seen at 0.8-1.0 um. Improving our Kurucz models for non-solar abundance and/or rapid stellar rotation we think is likely to remove this small discrepancy, and we currently trust an F2V assignment based on the strengths of various NIR absorption lines best. Our assignment is close to, but slightly different, than the F3 IV/V classification reported by Boyajian *et al.* 2015.

Our observations are consistent with a normal main sequence star without sufficiently large amounts of circumstellar gas or dust, or inflowing or outflowing material, to produce a SpeX detection via the scattering or thermal re-radiation of the star's insolation. Our observations have eliminated the possibility that there is a sizeable amount of very close-in, very hot (>1000 K) material along our line of sight to the star during the IRTF observations on 31 Oct 2015. Such a population of hot dust could obscure a large solid angle for a highly inclined, nearly edge-on system, and account for the up to 20% drops in the Kepler lightcurve flux, and was not previously ruled out, as obscuration by asteroidal or Kuiper Belt dust was, by the high resolution optical spectroscopy and multi-wavelength photometry presented in Boyajian *et al.* (2015). We can further state that there are none of the hallmarks of a young, pre-MS system for KIC8462852. It lacks the usual "flat", non-photospheric 2-4 um flux continuum and the gas emission lines seen in other YSO spectra from our NIRDS survey, like HD 51 Oph, HD 166191, or HD 165014.

Presuming we observed the system during its quiescent, or non-obscured, phase (a more than 99% probability, assuming the two major Kepler lightcurve dips reported by Boyajian et al. (2015) were 3 days wide and repeat every 750 days), any model for this system's strongly varying Kepler lightcurve will have to meet these baseline constraints on material outside the stellar photosphere. I.e., the source of the obscuring material must be episodic in nature, with a localized source reservoir and sink, surrounding a mature, Gyr-old star. Otherwise the material would have been detected by re-radiation of intercepted starlight in the infrared wavelengths. [Our observations cannot, however, rule out any artificial structure with an abnormally low (<1%) emissivity pointed in our direction.]

The cometary infall and breakup model most favored by Boyajian *et al.* 2015 fits our new constraints well. Considering the minimum amount of dust mass required to obscure 20% of a star's light, though, requires the conversion of a very large number of 1 km-radius comets ($\sim 10^5$ or more) every 750 days into ephemeral 1 μm dust particles, the smearing out of $>10^5$ 1 km-radius comets into > 100 μm radiation pressure stable heavy dust grains along an orbital trajectory, or the violent outgassing of one very big comet-like body of > 100 km radius (aka as a Centaur or KBO in our solar system) passing near the star's corona. These scenarios are interesting, in that we have been studying for some time now (using cold and warm Spitzer and SpeX) another mature, early-type F-star with many similarities, the F2V η Corvi system (Lisse *et al.* 2012, Lisse, Marengo *et al.* 2016 in preparation). In 2005 Wyatt *et al.* drew the community's attention to this system, demonstrating how it had both a bright, inclined (~45º to the line-of-sight) disk shaped Kuiper Belt to go along with its IRAS excess; in 2006 Chen *et al.* showed that the system had a very interesting mid-IR excess spectrum as well due to an inner belt of warmish dust at ~350K. Our 2012 follow-up work demonstrated a spectral connection between the warm dust and the cometary and KBO material in the system, and posited the formative mechanism as the infall from the Kuiper Belt of primitive icy bodies (either > $10^4$ comets/yr or 1 large KBO of > 130 km radius) due to an ongoing Late Heavy Bombardment. But unlike KIC8462852, η Corvi also demonstrates an NIR excess flux above the stellar photosphere of a few % (Fig 1), and thus an inner belt or ring of permanent circumstellar material. Could these two systems be both undergoing Late Heavy Bombardments, but the one in KIC8462852 is just starting?

## 5. Acknowledgements

The SPEX data used in this work was obtained by the authors as Visiting Astronomers at the Infrared Telescope Facility, which is operated by the University of Hawaii under contract with the National Aeronautics and Space Administration, Science Mission Directorate, Planetary Astronomy Program. Our NIRDS observations take advantage of and add to the SPEX spectral library of ~200 cool FGKM stars (Rayner *et al.* 2009), and we are deeply indebted to J. Rayner for providing his time and expertise to this project. The authors also wish to recognize and acknowledge the very significant cultural role and reverence that the summit of Mauna Kea has always had within the indigenous Hawaiian community. We are most fortunate to have the opportunity to conduct observations from this mountain.